\documentstyle[12pt]{article}
\newcommand{\be}{\begin{equation}}
\newcommand{\ee}{\end{equation}}
\newcommand{\s}{\section}
\newcommand{\ci}{\cite}
\newcommand{\r}{\ref}
\begin{document}
\begin{titlepage}

\begin{center}

\vskip 2cm
{\Large {\bf Dynamical confinement in bosonized $QCD_2$ $^*$}}\\

\vskip 1.7cm

{\large {A. Ferrando$^{\dag}$}}

\vskip 0.2cm
{Institute for Theoretical Physics}\\
{University of Bern}\\
{Sidlerstrasse 5, CH-3012 Bern, Switzerland}\\
\vskip 0.2cm
and

\vskip 0.3cm

{\large { V.Vento $^{\ddag}$}}

\vskip 0.2cm
{Departament de F\'{\i}sica Te\`{o}rica and I.F.I.C.}\\
{Centre Mixt Universitat de Val\`{e}ncia -- C.S.I.C.}\\
{E-46100 Burjassot (Val\`{e}ncia), Spain.}

\vspace{1.8cm}
{\bf Abstract}
\vspace{0.2cm}

\begin{quotation}
{\small In the bosonized version of two dimensional theories non trivial
boundary conditions (topology) play a crucial role. They are inevitable
if one wants to describe non singlet states. In abelian bosonization,
color is the charge of a topological current in terms of a non-linear
meson field. We show that confinement appears as the dynamical collapse
of the topology associated with its non trivial boundary conditions.}
\end{quotation}

\end{center}
\vspace{2.5cm}
$^*$Supported in part by CICYT grant \# AEN90-0040 and DGICYT
grant \# PB88-0064. \\
$^{\dag}$ Post Doctoral Fellow of CICYT- Plan Nacional de Altas
Energ\'{\i}as. \\
$^{\ddag}$ Vento @ Evalun11 ; Vento@vm.ci.uv.es
\end{titlepage}
\baselineskip  0.2in

\s{Introduction}
Quantum Chromodynamics in two dimensions is a confining theory.
The realization is quite simple, as can be seen from the
propagator, a solution to the non-dynamical gluon
equation,
\be
\partial^{-2}_- (x,y) = \frac{1}{2} |x_+ - y_+| \delta (x_- - y_-)
\ee
and which to first order in $\frac{1}{N}$ is proportional to the
$q\bar{q}$ interaction. Thus confinement here is a peculiarity of the
dimensionality of space-time \ci{co85}. The resolution of 't
Hooft's equation confirms this mechanism leading to a discrete,
stable and infinite spectrum \ci{ho74}. However the formalism is
much more powerful eliminating those amplitudes which would
violate confinement explicitly \ci{cc76,ei76}. Let us study for example
the process $meson \rightarrow q \bar{q} $. It can be shown that the
amplitude for it is given by
\be
F^{a \bar{b}}_n(t,r_-) = \frac{e^2}{r_-\sqrt{\pi N}} {\em
P}\int^1_0 dt' \frac{\varphi^{a  \bar{b}}_n (t')}{(t - t')^2}
\label{meson}
\ee
if $t \in [0,1]$. The notation follows that of ref.(\ci{ho74}). In
order to obtain the physical amplitude one has to impose the on
mass shell restrictions, i.e.,
\be
p^2 = M_a^2 ,\;\; \;\;\; (p-r)^2 = M_{\bar{b}}^2 ,
\;\;\;\;\; r^2 = r^n
\ee
where $M_i$ are the renormalized quark masses and $r_n$ the
corresponding meson mass. The on mass shell condition leads to
the following relation for the adimensional momentum $t$
\be
\mu ^2 = \frac{\alpha_a}{t} + \frac{\alpha_{\bar{b}}}{1-t}
\ee
and therefore 't Hooft's equation becomes
\be
{\em P} \int^1_0 \frac{\varphi^{a \bar{b}_n}}{(t - t')^2} = 0
\ee
which implies the vanishing of the amplitude for the process.
Therefore an expected consequence of the
confinement mechanism is that no quark can be liberated from a
bound state.

Let us characterize confinement from the point of view of the
quarks. 't Hooft \ci{ho74} used a cut-off to regulate the
infrared divergences. In this way the quark propagator acquired a
singularity leading to an infinite renormalized mass. This
behavior was interpreted as the confinement mechanism. The
pole of the quark propagator was displaced from the bare quark
mass to infinity leading to an impossibility of satisfying the
on mass shell condition. This argument was heavily critized once
it was discovered that the choosing of the infrared regulator is
just a consequence of gauge freedom \ci{ei76}. One can choose
a gauge where the self-energy is perfectly finite. However we
next show that one can find a relation between the self-energy
and the interaction potential and in this way the latter might be
used to describe confinement.

Dyson's equation for the self-energy is
\be
\Sigma (t) = -\frac{e^2 r_-}{2 \pi} {\em P}
\int^{\infty}_{-\infty} \frac{sign(t')}{(t-t')^2}
\ee
which in position space leads to
\be
\Sigma (x_+) \sim \frac{V(x_+)}{x_+}
\ee
Thus the long distance behavior of the self-energy contains
information about the confining features of the potential.

It is important to stress that
higher order terms in $\frac{1}{N}$ do not change the confining
character of the interaction. In particular, in the so-called
singular gauge \ci{cc76}, it can be shown that the higher order
corrections are proportional to the gauge parameter $\alpha$
\be
V(x_+) \sim |x_+| -\alpha (1 + f(\frac{1}{N}))
\ee
and therefore do not affect the result of this calculation.

The confinement mechanism just described is not a
{\em capriccio} of 't Hooft's model but a fundamental property of
$QCD_2$ \ci{bo84,hn88}.

Two dimensional theories are exactly bosonizable and we have seen
that in the chiral limit many crucial properties of the spectrum
and the realization of chiral symmetry appear in a clear and
appealing fashion \ci{fv91}. The purpose of this paper is to
analyze confinement in this description of the theory. The
crucial symmetry associated with confinement is color. For reasons
of simplicity we shall work only with one flavor and two colors,
a description which allows visualization of the results but does
not imply any restriction on the physics involved.

\s{Realization of color symmetry in the bosonized {\em free} massless
theory}

The bosonized action for a massless quark using abelian
bosonization is \ci{ha75}
\be
{\em S} = \int d^2x \{ \frac{1}{2}(\partial_{\mu} \varphi)^2 +
\frac{1}{2}(\partial_{\mu} \eta)^2 \}
\label{action}
\ee
It can seem surprising that such a simple action can realize the
full global $SU(2)$ color symmetry. A mere counting of color
degrees of freedom seems to imply that some color field might be
missing \footnote{Non-abelian bosonization would lead to an
action in terms of the $SU(2)$ fields $g = \exp{(i\frac{\tau^a
\pi^a}{2})}, a=1, 2, 3,$ with three color degrees of freedom, where the
$\pi^a$ fields transform according to the adjoint representation.}.
The answer to this apparent contradiction lies in the {\em
topology} of the $\eta$ field. This field does not transform
under any linear transformation of SU(2), but it can be related
through the Mandelstam equations to the $s^{\alpha}_{\pm}$
operators, which transforms under the fundamental representation
of the color group \ci{ha75}. But in order for this relation to
occur the $\eta$ field must possess non-trivial boundary
conditions, i.e.,
\be
T^3 = \frac{1}{\sqrt{2\pi}}[\eta (+\infty) - \eta (-\infty)]
\label{t3}
\ee
Thus the action in Eq.(\r{action}) can only realize the full
$SU(2)$ group if one incorporates besides the conventional
solutions $\eta _0
(\pm \infty) = 0 $, those associated with non-trivial boundary
conditions ($\eta(+\infty) = \pm \sqrt{\frac{\pi}{2}}\; , \;\;\;
\eta(-\infty) = 0$). That they exist for
the free theory has been proven by construction \ci{ha75}. Moreover they
are dynamically allowed since
the finiteness of the energy for physical states,
\be
{\em E}_{\eta} = \int dx {\em H}(\eta) = \int dx \frac{1}{2}
(\partial _x \eta)^2 < \infty
\ee
only requires the asymptotic vanishing of the derivative. Thus
the $\eta$ field can tend to different constants at $\pm \infty$,
and generate solitonic solutions. These states are stable because
they are protected from desintegrating into conventional $T^3 = 0$
states by topological conservation laws.

Despite its naive appearence, the bosonized action
Eq.(\r{action}) has an internal symmetry group which is bigger
than the obvious $U(1)_F \otimes U(1)_C$. The reason behind this
statement is that the soliton operator appearing in Eq.(\r{t3})
generates besides the conserved current
\be
J^3_{\mu} = \frac{1}{2} \varepsilon_{\mu \nu} \partial^{\nu} \eta
\ee
two others, $J^1_{\mu}$ and $J^2_{\mu}$, whose explicit
expresions are highly non-local \ci{ha75}. The main difference
between them is that for the former the conservation law is a
trivial mathematical identity (topological conservation law)
while for the latters the dynamics of the $\eta$ field is
required for conservation (Noether currents). The conserved
charges associated to these currents close an $SU(2)$ algebra.
Moreover the following commutator can be calculated
\be
[T^a, s^{\alpha}_{\pm}] = \left( \frac{\tau^a}{2}\right)_{\alpha \beta}
s^{\beta}_{\pm}
\ee
which indicates that the soliton operators transform under the
color group in the fundamental representation.

It is important to stress that it is the existence of non-trivial
boundary conditions which generate the full $SU(2)_C$ algebra.
The raising and lowering operators connect different topological
sectors, i.e.,
\be
T_- |\eta (\infty) = \sqrt{2\pi} t> \propto
|\eta (\infty) = \sqrt{2\pi} (t-1)>
\ee
Therefore if one would restricts by {\em fiat} the Fock space to
that determined by ordinary boundary conditions (obc's), one is
eliminating without justification all states with $T_3 \neq 0$.

\s{Vacuum structure of the bosonized {\em free} massless theory}

As we have just seen well-defined non-trivial boundary conditions
in the bosonized theory are essential to reach the full $SU(2)_C$
symmetry of the fermionic action. These boundary conditions are
not explicit in the bosonized action. However it is possible to
introduced them by adding a small mass term into the fermionic
action. The bosonized form of this mass term becomes
\be
{\em S_M} = \int d^2x \{ \mu M (\cos{\sqrt{2\pi} \varphi})_{\mu}
(\cos{\sqrt{2\pi} \eta})_{\mu}\}
\ee
$\mu$ is a renormalization mass \ci{co75}. The classical potential
has a minimum at
\be
{\em V_M}(\tilde{\varphi},\tilde{\eta}) = - \mu M
\ee
which implies a degenerate vacuum formed by the infinite set of
points \ci{ha75}
\be
\emptyset = \emptyset_I \cup \emptyset_{II}
\ee
where
\[
\emptyset \equiv \left\{ \begin{array}{llr}
(\sqrt{2\pi}n, \sqrt{2\pi} m) &  \in I &\\
(\sqrt{2\pi}(n +\frac{1}{2}), \sqrt{2\pi}(m + \frac{1}{2})) & \in
II &\\
 & & (n,m) \in {\em Z}
  \end{array} \right. \]
\be
\label{vacuum}
\ee
The structure of the vacuum gives us precise information about
the solutions of the bosonized theory. We have already shown how
to read particle properties from field boundary conditions . When
a potential is present, the possible boundary conditions are
related to the minima of the potential \ci{co85,cl84}
\be
\lim_{x \rightarrow \pm \infty} (\varphi(x), \eta(x)) =
(\varphi_{n_{\pm}}, \eta_{m_{\pm}}) \in \emptyset
\label{boundary}
\ee \\
Baryon number and $T^3$ charge are related by the bosonization
formulae to the asymptotic conditions of the fields
\begin{eqnarray}
B & = & \frac{1}{\sqrt{2\pi}}[\varphi (+\infty) - \varphi
(-\infty)]  \nonumber \\
T^3 & = & \frac{1}{\sqrt{2\pi}}[\eta (+\infty) - \eta
(-\infty)]  \nonumber \\
\label{charges}
\end{eqnarray}
Eqs.(\r{boundary}) and (\r{charges}) allow us to transform
the $(\varphi,\eta)$ vacuum structure plane, Eq.(\r{vacuum}),
into a diagram for physical states.
%************************************************************
% Figure 1
\begin{figure}
\vspace{3.0in}
\caption{Examples of allowed states: a) Diagonal arrows: up-color
quark state ($B = \frac{1}{2} , T^3 = \frac{1}{2}$); b) Horizontal
arrows: up-color vector state ($B = 0 , T^3 = 1$); c) Vertical
arrows: color singlet baryon state ($B= 1, T^3 = 0$).}
\label{boundaries}
\end{figure}

Fig.(\r{boundaries}) shows that all states described by
arrows of the same length and direction are equivalent, i.e.,
they represent the same physical state. Thus we can define an
equivalence relation and choose just one representative per
class. For example, we proceed by attaching the arrows to the same
point ($\varphi (-\infty) =  \eta (-\infty) = 0$), since we have the
freedom to select one of the boundary conditions. With this
restriction all the physical states will be represented by the
points of the vacuum structure lattice, which
becomes in this way a $(B,T^3)$ plane (see Fig.(\r{states}))
%*************************************************************
%Figure 2
\begin{figure}
\vspace{3.0in}
\caption{Representation of the physical states in the $(B\; ,\; T^3)$
plane.}
\label{states}
\end{figure}

Global $SU(2)$ transformations leave the potential invariant.
This is easily seen in the non-abelian bosonization scheme where
the potential is proportional to $tr(g) = \cos{\sqrt{2\pi}\eta}$,
an $SU(2)$ invariant, and a function of $\varphi$, itself invariant by
construction \ci{fv91}. Therefore the set of physical states, the
lattice, is invariant under these transformations. Moreover it is
also invariant under the discrete translations \ci{ha75}
\[\begin{array}{cccr}
\varphi & \rightarrow & \varphi + \sqrt{2\pi} n &\\
\eta & \rightarrow & \eta + \sqrt{2\pi} m &\\
    & & & (n, m) \in  {\em Z} \\
\end{array} \]
\be
\ee

Before closing this section it is important to emphasize that the
vacuum structure just studied coincides with that of the chiral
limit of the theory. The topological charges of the physical
states are mass independent. They survive in the chiral limit
together with the asymptotic conditions that generate them. The
minima of the potential become the non-trivial boundary
conditions, which are allowed by energy considerations.

\s{The vacuum structure of $QCD_2$}
Our starting point will be Baluni's bosonized action \ci{ba80}
\footnote{There are indications that even in the non-abelian case it is
possible to find a similar form for the bosonized
action\ci{ei76}.}. In this case the potential arising from the
quark-gluon interaction is
\be
{\em V_i}(\eta) = \frac{e^2}{32 \pi} \eta^2 +
\sqrt{\pi} \Lambda^2 \{ 1 - \frac{\sin{\sqrt{2\pi}
\eta}}{\sqrt{2\pi} \eta} \}
\label{potential}
\ee
Here $\Lambda$ is a scale parameter coming from the
renormalization through normal-ordering.

Let us generalize our energy finiteness argument to the
interacting case
\be
{\em E}_{\eta} = \int^{+\infty}_{-\infty} dx [\frac{1}{2}
(\partial_x \eta)^2 + {\em V_i}(\eta)]
\ee
A soliton field $\eta_s$ must satisfy the equations of motion.
For static solutions we have
\be
{\em E}_{\eta_s} = \int^{\infty}_{-\infty} dx \; 2{\em V_i}(\eta_s)
\label{potenergy}
\ee
But this integral is necessarily divergent unless $\eta_s (+\infty)$ is an
absolute minimum \footnote{ In a recent paper by Ellis et al.
\ci{ef91} they find color solitons in bosonized $QCD_2$ with
infinite energy. We claim that their {\em constituent quarks}
correspond to solitons which do not connect absolute minima.}.
This clearly implies that states which are not constructed at
an absolute minimum have infinite energy.

The potential Eq.(\r{potential}) is positive definite and it
has only one absolute minimum  ${\em V_i}(\eta) = 0$ which occurs
for $\eta = 0$
\be
{\em V_i}(\eta) = 0 \leftrightarrow \eta = 0
\ee
This property is crucial to understand the vacuum structure of
bosonized $QCD_2$. Even if a mass term was added, the same result
would persist. What are the implications of this result on the
$(B, T^3)$ plane?

The full potential has a lower bound
\be
{\em V} (\varphi, \eta) =  {\em V_M}(\varphi, \eta) + {\em V_i}
(\eta) \geq -\mu M , \;\; \forall (\varphi,\eta)
\ee
The equality is only saturated if the following conditions are
met
\be
\cos{\sqrt{2\pi}\varphi} \cos{\sqrt{2\pi}\eta} = 1 \;\;\; and
\;\;\; {\em V_i({\eta})} = 0
\ee
Thus the minima are the set of points
\be
\emptyset_{QDC_2} = \{(\varphi_n ,\; \eta = 0) \;\; ; n\in {\em Z} \}
\ee
which correspond to the color singlet states shown in
Fig.(\r{singlet}).
%************************************************************
%Figure 3
\begin{figure}
\vspace{3.0in}
\caption{ The color singlet sector of $QCD_2$.}
\label{singlet}
\end{figure}

Let us expand on this subtle point. Dynamically the topologically
non-trivial solutions are pushed to infinite energy and thus the Fock
space in $QCD_2$ has been reduced to that of only $T_3 = 0$
states. We claim however that in this case this corresponds to
only the color singlet states, i.e., $T = 0$ states. Let us prove
it.

In order to do so we shall calculate the matrix element of
$\vec{T} \cdot \vec{T}$ between any such states,
\be
<\Phi| \vec{T} \cdot \vec{T}|\Phi'> = <\Phi|T_3 T_3 +
\frac{1}{2} T_-T_+ + \frac{1}{2}T_+T_-|\Phi'>
\ee
where evidently these charge operators are obtainable from the
space integrals of the zeroth component of the color currents
$J^{\mu}_a$ and the states belong to the Fock space determined
by obc's. In order to calculate the matrix elements of the ladder
operators, we use completeness in this space
\be
<\Phi|T_{\pm}T_{\mp}|\Phi'> = \sum_n <\Phi|T_{\pm}|\Phi_n>
<\Phi_n|T_{\mp}|\Phi'>
\ee
The result limited to this subspace is trivially zero and thus
the states are necessarily color singlet. Note that the
restriction in the present case has not been imposed by {\em
fiat} but has been dynamically generated and therefore is a
consequence of the interactions in $QCD_2$. States with non vanishing
color quantum numbers acquire infinite mass as a result of
eq.(\ref{potenergy}).

The presence of the color interaction has transformed completely
the structure of the vacuum with respect to that of the {\em free}
theory. Once we turn on the interaction ($e \neq 0$), the {\em free}
vacuum collapses immediatly to the bosonized $QCD_2$ vacuum and
therefore the set of all possible physical states reduces to the
subset of color singlet states. Confinement appears in this picture
as a collapse of topology, resulting from the dynamical
prohibition of the non-trivial boundary conditions for the $\eta$
field.

To conclude, the singlet spectrum is formed by particles of
baryon number $B = n$, i.e., by mesons ($n = 0$) and by baryons
($ n = 1, 2, ...$) or antibaryons ($n = -1, -2, ...$). Particles
with half-integer baryon number are {\em necessarily} colored
states. This result agrees with what is expected from the
fermionic theory. Color singlet states are operators which in
color space are of the form
\be
q^+_{\alpha} q_{\alpha}\;\; , \;\;\; \varepsilon^{\alpha \beta}q_{\alpha}
q_{\beta}\;\; , \;\;\; q^+_{\alpha} q_{\alpha} \varepsilon^{\beta \gamma}
q_{\beta} q_{\gamma}\;\; , \;\;\; \ldots
\ee

\s{Confinement and topology}

Confinement is recognized in the effective action by the fact
that the  $\eta$ field is a color singlet field. No color charged
states can be built from the bosonized action. As we have proven
in previous work \ci{fv91}, resonant states belong to the color
sector of the bosonized theory. Consequently the $\eta$ dynamics
is responsible for generating the spectrum and therefore only
color singlet states can appear.

Since $\eta$ is a color singlet it must have simple relations
with $SU(2)_C$ invariant quantities. This is easily established
when we compare simple operators in their abelian and non-abelian
forms. For example, we can express the non-abelian bosonized
field $g$ in its abelian form
\be
trg = g^1_1 + g^2_2 = 2\cos{\sqrt{2\pi}\eta}
\ee \\
But because $g \in SU(2)_C$ it also has a standard representation
in terms of the adjoint fields $\pi^a$, $g^{\alpha}_{\beta} =
e^{i\pi^a\left(\frac{\tau^a}{2}\right)_{\alpha \beta}} $ and therefore $\eta =
\frac{\sqrt{\pi^a \pi_a}}{2\sqrt{2\pi}}$. Thus for ordinary
boundary conditions the $\eta$ field is a color singlet, moreover
it is a singlet scalar particle.

We have learned that the vacuum structure of the {\em free} field
theory is the consequence of the existence of non-trivial
boundary conditions. The stability of the soliton numbers
$(B,T^3)$ is guaranteed by the existence of {\em topological}
conservation laws for the $U(1)_F$ and $U(1)_C$ currents
\be
j^{\mu} = \frac{1}{\sqrt{2\pi}} \varepsilon^{\mu
\nu}\partial_{\nu} \varphi \;\; , \;\;\;
J_3^{\mu} = \frac{1}{\sqrt{2\pi}} \varepsilon^{\mu
\nu}\partial_{\nu} \eta
\ee
Because the $(B,T^3)$ charges depend on the values of the
$(\varphi,\eta)$ fields on the border of the one-dimensional
space, they can be related to topological properties of the
groups associated with them. The $(B, T^3)$ charges generate the
$U(1)_F \otimes U(1)_C$ explicit symmetry of the {\em free}
lagrangian. Its action can be written in terms of the
$U(1)_F \otimes U(1)_C$  fields  $(e^{i\sqrt{2\pi}\varphi},
e^{i\sqrt{2\pi}\eta})$ using
\be
\partial_{\mu} \phi \partial^{\mu} \phi = \frac{1}{2\pi}
\partial_{\mu}(e^{i\sqrt{2\pi}\phi})
\partial^{\mu} e^{-i\sqrt{2\pi} \phi}
\ee
where $\phi = \varphi, \;\eta$.

The compactified one-dimensional space is the ${\em S^1}$ sphere.
We may define the {\em free} action in terms of the
$U(1)_F \otimes U(1)_C$  fields with support on the sphere. These
fields map the sphere into the group, and since $U(1)$
has a parameter space which is the unit circle, it turns out to
be a mapping from $S^1$ into $S^1\otimes S^1$. However this
mapping is not unambiguously defined, and each of the homotopic
solutions can be charaterized by two integers $(\nu_F,\nu_C)$.
These topological charges can be obtained by means of the
integral formulas in terms of group elements \ci{co85}
\be
\nu_i = \frac{i}{2\pi} \int^{2\pi}_0 d\theta g_i
\frac{d}{d\theta} g_i^{-1}
\ee
where $g_i \in U_i(1), i = F, C$ and therefore
\[\begin{array}{ccccc}
\nu_F & = & \sqrt{\frac{2}{\pi}}[\varphi (2\pi) - \varphi (0)] & =
& 2B \\
\nu_C & = & \sqrt{\frac{2}{\pi}}[\eta (2\pi) - \eta (0)] & =
& 2T^3 \\
\end{array} \]
\be
\ee
The non-trivial boundary conditions produce the {\em winding
numbers} associated with the homotopy classes of these mappings.
The lattice of physical states gives just the first homotopy
class of the group
\be
\emptyset \approx \Pi_1[U(1)_F \otimes U(1)_C] \approx \Pi_1[U(1)_F
\times \ \Pi_1[U(1)_C] \approx {\em Z_F} \otimes {\em Z_C}
\label{topol}
\ee
Let us add to the {\em free} action a non-confining interaction,
then the quarks must be the only asymptotic in/out states. Thus
asymptotically the action is that of a free $(\varphi,\eta)$
field with non-trivial boundary conditions and therefore we
obtain a free vacuum structure. Due to the conservation of the
charges $(B,T^3)$ this structure is also that of the interaction
theory, because if not, asymptotic states different from free
quark states could be generated, contrary to our starting
assumption. Thus the addition of a non-confining interaction
leaves the vacuum structure unaltered and therefore we can
characterize this type of interactions topologically by
Eq.(\r{topol}).

Let us now add a confining interaction. The vacuum structure
collapses into a colorless subset (see Fig.(\r{singlet})). In
topological language the non-trivial topology induced by the
mapping
\be
\Pi_1[U(1)_C] \approx {\em Z_C}
\ee \\
disappears. The confining interaction forces the physical
solutions to be states of $T^3 = \nu_C = 0$. The color topology
becomes trivial and therefore the topology induced by the
interaction in the center of the $SU(2)_C$ gauge group ($U(1)_C$)
determines that all physical states have to be color singlets.

\s{Conclusion}

The fermionic description indicates in a qualitative manner that
$QCD_2$ is a confining theory beyond leading order
in the $\frac{1}{N}$ expansion. The spectrum,
as well as, the vanishing of the quark creation amplitudes, corroborate
the non existence of {\em free} color non singlet states. The quark
self-energy shows a confining behavior.

In the bosonized version of fermionic two dimensional theories
topology plays an important role. States with baryon number and
color charge are described by solitons. The properties of the vacuum,
which give rise to non-trivial boundary conditions, determine the quantum
number structure of the Fock space. We have analyzed initially the rich
spectrum  of non confining theories by discussing the role of the
boundary conditions. It is appealing that in the bosonized version of the
theory , this discussion can be carried out purely at the classical level.

Once the dynamics of $QCD_2$ is incorporated the Fock space collapses
to the subset of color singlet states and we recover the fermionic
result independent of any large $N$ assumption. The mechanism
can be cast in a more mathematical language by invoquing homotopy
groups, but one should not avoid the very naive dynamical statement,
namely that color solitons are given infinite energy \ci{ef91}.

The abstraction associated with the mathematical language might guide one
into the four dimensional case. Is it not possible to write  an approximate
bosonized theory in four dimensions? Skyrme type models have been extremely
succesful in describing the low energy flavor properties of the theory , but
they avoid confinement simply by assuming its existence. Could one find an
approximate bosonized lagrangian in terms of the color degrees of freedom ?,
and if so, what is the mechanism that produces the collapse of the spectrum ?
The work of 't Hooft \ci{ho80} has been pioneering in this respect, but again
the dimensionality of space-time makes difficult the connection. No relation
between our topological scheme and his has been as of yet found, but the
endeavor seems sufficiently appealing to embark.

\s*{Acknowledgement}
We have received useful comments and criticism from M. Asorey, A.
Gonz\'alez-Arroyo, D. Espriu, J. Ros, A. Santamar\'{\i}a and J.Segura. One
of us (A.F.) is grateful to H. Leutwyler, P. Minkowski, A.
Smilga and U. Wiese for illuminating discussions.

\end{document}